\documentclass [epsfig,twocolumn,pra,showpacs,preprintnumbers,amsmath,amssymb,showkeys]{revtex4}
\usepackage{graphicx}

\begin{document}
\bibliographystyle{apsrev}

\title{Hawking radiation in sonic black holes}
\author{S. Giovanazzi}
\email{sg51@st-andrews.ac.uk} 
\affiliation{School of Physics and Astronomy, University of St Andrews,
North Haugh, St Andrews, Fife, KY 16 9SS, Scotland}
\date{\today}

\begin{abstract}
I present a microscopic description of Hawking radiation in sonic black holes. A one-dimensional Fermi-degenerate liquid squeezed by a smooth barrier forms a transonic flow, a sonic analogue of a black hole. The quantum treatment of the non-interacting case establishes a close relationship between the Hawking radiation and quantum tunnelling through the barrier. Quasi-particle excitations appear at the barrier and are then radiated with a thermal distribution in exact agreement with Hawking's formula. The signature of the radiation can be found in the dynamic structure factor, which can be measured in a scattering experiment. The possibility for experimental verification of this new transport phenomenon for ultra-cold atoms is discussed.
\end{abstract}
\maketitle

\emph{Introduction.} --- One of the most fascinating effects in astrophysics is Hawking's prediction of black hole radiation \cite{Hawking74}. Hawking realized that particles such as neutrinos and photons could be created spontaneously from any black hole and emitted at just the rate that one would expect if the black hole is a body with a temperature $\hbar \kappa / 2\pi k_B$ proportional to its surface gravity $\kappa$. Although Hawking's prediction combines gravitation and quantum mechanics to produce thermodynamics, it was rapidly appreciated that the radiation predicted by Hawking is more primitive and fundamental. A beautiful example of this was pointed out by Unruh \cite{Unruh81}: The propagation of sound in a fluid/gas turning supersonic, such as in the nozzle of a rocket engine, is similar to the propagation of a scalar field close to a black hole. By adapting Hawking's derivation, Unruh predicted the analogue of Hawking radiation for sound waves in a moving fluid. 
This sonic analogue could provide a way to experimentally test Hawking's effect.

Although Unruh's prediction is not restricted a priori to a specific fluid, superfluids immediately were considered as the natural candidates.
Superfluid helium II has been the first system to be proposed for the investigation of the Hawking effect \cite{BookABH}. More recently atomic Bose-Einstein condensates have been considered, as they may offer a better chance of testing Hawking radiation  \cite{Visser98,Garay,Fischer}. 
Transonic flows could be also realized in one-dimensional (1D) systems \cite{Garay,Giovanazzi04} and available 1D quantum systems, including ultra-cold atoms in optical potentials and wave-guides \cite{Goerlitz01}, could be adapted to model a sonic black hole analogue and to investigate the elusive Hawking effect.

The Hawking effect has not been considered so far in systems of non-interacting fermions. 
It is the aim of this Letter to show that indeed a 1D Fermi-degenerate non-interacting gas that scatters against a very smooth potential barrier provides a clear and straightforward quantum mechanical microscopic description of Hawking radiation.
Sonic Hawking radiation is formed due to quantum tunnelling through the top of the potential barrier. Particle-hole excitations are created at the barrier and are then radiated with a momentum distribution characterized by a temperature in exact agreement with Hawking-Unruh's formula \cite{Hawking74,Unruh81,Visser98}. Although the evolution of the fluid's flow is unitary, when the momentum distribution is measured locally, for instance through the measurement of the dynamic structure factor, it is not distinguishable from a thermal (and incoherent) distribution. 
The general case of 1D quantum fluids as well as the possible relevance for experiments is also discussed.

\emph{Part I. Microscopic description.} --- 
Consider a Fermi-degenerate gas of particles moving in a sufficiently narrow and smooth constriction such that their motion can be considered 1D.
Their stationary single-particle wave-functions 
can be written in the adiabatic approximation \cite{Glazman88} as $\Psi(x,y,z)=\psi(x)\xi_{0;x}(y,z)$ where $\xi_{0;x}(y,z)$ is the transverse ground-state wave-function with eigenenergy $\epsilon_{0}(x)$. $\psi(x)$ satisfies a 1D Schr\"{o}dinger equation $[-(\hbar^{2}/2m)\partial_{x}^{2}+ V_{\text{ext}}(x)]\psi(x)=\epsilon\psi(x)$ where $V_{\text{ext}}$ is an effective potential that includes also the zero point energy $\epsilon_{0}(x)$ \cite{eff}. I assume $V_{\text{ext}}$ very smooth and essentially non-zero only near its maximum $V_{\text{max}}$ located at $x=0$.

Consider a flow of particles coming from the left and colliding against the potential $V_{\text{ext}}$.
Their particle's wave-functions are given by scattering wave functions asymptotically defined by  $\psi_{\epsilon} \propto  [ e^{i k x} + r(\epsilon) e^{- i k x} ] \Theta(-x)+ t(\epsilon) e^{i k x}  \Theta(x) $, where $\epsilon=\hbar^{2}k^{2}/2m$ is the energy eigenvalue and $r(\epsilon)$ and $t(\epsilon)$  are the reflection and transmission amplitudes, respectively.
The evolution of the particle flow is unitary and the many-body wave function of the flow of particles is represented in a second quantization by
\begin{equation}
| f \rangle = \prod_{0 \le \epsilon < \epsilon_{\text{max}}} a^{\dag}_{\epsilon}| - \rangle
\label{flow}
\end{equation}
where $a^{\dag}_{\epsilon}$ is the creation operator of $\psi_{\epsilon}$'s and $| - \rangle$ is the vaccum of particles. In this state all $\psi_{\epsilon}$'s are occupied up to an energy $\epsilon_{\text{max}}$, which is assumed higher than $V_{\text{max}}$.
The non-interacting case is an ideal example of a fluid turning supersonic (see below) where the microscopic wave function of the flow (\ref{flow}) is given from the beginning.

Since the barrier potential is smooth the particle propagation is semi classical and in first approximation $|t(\epsilon)|^{2} \approx \Theta(\epsilon-V_{\text{max}})$.
Therefore the fluid maintains in a first approximation the sharp Fermi-degenerate character of the local momentum (or velocity) distribution, which represents the ``Vacuum'' of the black hole sonic analogue.
This allows us to describe the fluid dynamic with the standard hydrodynamic equations.
The velocities of the particles in the fluid are uniformly distributed within an interval $(v_{L},v_{R})$. Figure 1 shows an example of velocity distribution. The right Fermi velocity $v_{R}=\sqrt{v_{\text{max}}^{2} - 2 V_{\text{ext}}(x)/m}$ is the velocity of a particle moving from the left to the right with initial velocity higher than the classical threshold velocity to go over the barrier $v_{\text{esc}}=\sqrt{2 V_{\text{max}}/m}$.
The left Fermi velocity $v_{L}$ is the velocity of a particle starting from the top of the barrier (located at $x=0$), i.e. $v_{L}= \text{sgn}(x) \; \sqrt{v_{\text{esc}}^{2} - 2 V_{\text{ext}}(x)/m}$. The $\text{sgn}(x)$ function accounts for the two possible different signs of the velocity that the particle can have falling either to the right or left of the barrier.
\begin{figure}[htbp]
\begin{center}
\includegraphics[width=85mm]{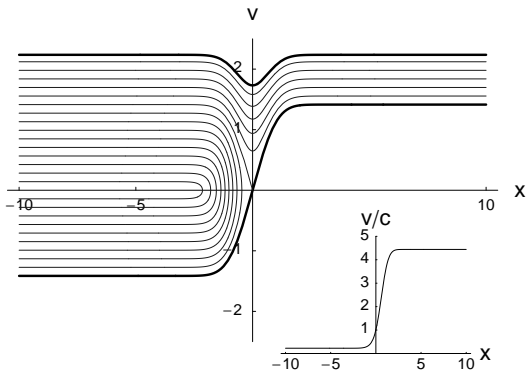}
\caption{Semi classical velocity distribution as function of the coordinate $x$. Velocities are uniformly distributed in the area limited by 
$v_{R}$ (top solid line) and $v_{L}$ (bottom solid line). Thinner solid lines correspond to different classical orbits of the particles of the fluid.
Here $\epsilon_{\text{max}} = 2.5$ and $V_{\text{ext}}  = \text{exp}(-x^{2})$ in dimensionless units. The inset shows the Mach number $v/c$ as function of position $x$.}
\label{fig1}
\end{center}
\end{figure}

The flow velocity $v=(v_{L}+v_{R})/2$ is the velocity of a local reference frame from where the fluid appears in equilibrium. Introducing the Fermi velocity $v_{F}=(-v_{\text{L}} + v_{\text{R}})/2$ as the Fermi velocity in the co-moving frame,  the density is given by $n=m v_{F} / \pi\hbar$. 
With these definitions the current $\Omega = n v=m v v_{F} / \pi\hbar$ is conserved \cite{notecurrent}. 
The Bernoulli equation $v^{2}/2 +  h + V_{\text{ext}}/m=\text{cost}$ where $h(n)=\pi^2\hbar^2 n^2/2m^{2}$ is
the enthalpy (the Fermi energy divided by the particle mass) is satisfied with $\text{cost}=v^{2}_{\text{esc}}/2+v^{2}_{\text{max}}/2$. 
The sound velocity ($c^2=n \,dh(n) / dn$) is consistently given by the Fermi velocity $c = v_{F}$. It is easy to verify that the above defined particles flow is a transonic flow, namely subsonic on the left of the barrier and supersonic on the right. The inset of Figure 1 shows the behaviour of the Mach number $v/c$ in the 1D channel.

Even if the potential barrier is very smooth, its finite thickness near its top smears the semi classical Fermi distributions at $v_{\text{L}}(x)$. This is due to those particles colliding against the barrier with energies comparable with $V_{\text{max}}$.
The reflection probability is given by (see also \cite{Hanggi90}) 
\begin{equation}
|r(\epsilon)|^{2}=\frac{1}{1+\exp(2\pi(\epsilon-V_{\text{max}})/\hbar\omega_{x})}
\label{penetration}
\end{equation}
where $\omega_{x}$ is the frequency of the inverted harmonic oscillator obtained expanding the barrier potential $V_{\text{ext}} = V_{\text{max}} - \frac 12 m \omega_{x}^2 x^2 + o(x^2)$ around its maximum. 
Hawking's formula for a sonic black hole \cite{Unruh81,Visser98} is given by
\begin{equation}
T_H=\left( \frac{\hbar}{2\pi k_B} \right) \frac{d(v-c)}{dx} \,,
\label{unruh}
\end{equation}
where the surface gravity $\kappa$ in Hawking temperature $\kappa\hbar / 2\pi k_B$ is replaced by 
$d(v-c)/dx$. For the transonic flow of fermions defined above $d(v-c)/dx=d(v_{L})/dx=\omega_{x}$ and Hawking temperature (\ref{unruh}) becomes
\begin{eqnarray}
T_{\text{H}}= \frac{\hbar \omega_{x}}{2\pi k_{\text{B}}}  \;.
\label{bare1}
\end{eqnarray}
Therefore the particles reflected from the barrier are distributed according to the reflection probability (\ref{penetration}), which is identical to a thermal Fermi distribution 
\begin{equation}
n_{T,\mu}(\epsilon)=\frac{1}{1+\exp((\epsilon-\mu)/k_{\text{B}}\,T)}
\label{thermaldistribution}
\end{equation}
where the chemical potential $\mu=V_{\text{max}}$ and the temperature is in exact agreement with Hawking temperature (\ref{bare1}).

Consider a probe beam intersecting our 1D flow in a region quite far from the barrier. The measurement of the differential scattering cross-sections with the probe beam allows obtaining the dynamic structure factor \cite{Pines}, which can be used to investigate Hawking radiation. 
The dynamic structure factor is defined as $S(q,\omega)=\Sigma_{l \ne f} \langle f | \rho_{q} | l \rangle \langle l | \rho_{q}^{\dag} | f \rangle  \delta (\omega-\omega_{l})$ where $\omega_{l}=(E_{l}-E_{f})/\hbar$ and $E_{l}$ are the excitation frequencies and the energies of the eigenstates $| l \rangle $, respectively. Positive (negative) frequencies in $S(q,\omega)$ are related to absorption (release) of energy from the system.
$\rho_{q}^{\dag} = \sum_{i} e^{i q (x_{i}-R_{p})} e^{-(x_{i}-R_{p})^{2}/L^{2}_{p}}$ is a density operator defined locally where $L_{p} \gg 1/q$. Here $L_{p}$ and $R_{p}$ are the size and the position of the region where the probe beam and the system beam interact, respectively \cite{note}.

The dynamic structure factor of the flow is given in the subsonic region by 
\begin{eqnarray}
S(q,\omega)=S^{\text{eq}}_{0,\mu_{R}}(q,\omega)\,\Theta\left(\frac{\omega}{q}\right)+S^{\text{eq}}_{T_{\text{H}},\mu_{L}}(q,\omega)\,\Theta\left(-\frac{\omega}{q}\right)
\end{eqnarray}
and in the supersonic region by 
\begin{eqnarray}
S(q,\omega)=\left[S^{\text{eq}}_{0,\mu_{R}}(q,\omega)+S^{\text{eq}}_{T_{\text{H}},\mu_{L}}(q,-\omega)\right]\Theta\left(\frac{\omega}{q}\right)
\end{eqnarray}
 where $\mu_{L}=V_{\text{max}}$ and $\mu_{R}=\epsilon_{\text{max}}$. 
$S^{\text{eq}}_{T,\mu}(q,\omega)$ is the dynamic structure factor of an equilibrium 1D Fermi gas at temperature $T$ and chemical potential $\mu$ given by $S^{\text{eq}}_{T,\mu}(q,\omega)= \frac{L_{\text{eff}}\, m}{4\pi\hbar^{2}q }\, e^{\hbar\omega/k_{\text{B}}T}/\{\text{Cosh}(\hbar\omega/k_{\text{B}}T)+
\text{Cosh}[(\frac{m\omega^{2}}{2q^{2}}+\frac{\hbar^{2} q^{2}}{8 m} -\mu)/k_{\text{B}}T]\}$
where $L_{\text{eff}}=\sqrt{\pi/2}L_{p} $ is an effective length of the probe-system interaction region \cite{note}. 
Its $T=0$ limit is given by 
$S^{eq}_{0,\mu}(q,\omega)=\frac{L_{\text{eff}}\, m}{2\pi\hbar^{2}q }\,\Theta\left(-\omega+\frac{\hbar^{2}q^{2}}{2m}+\frac{\hbar^{2}q k_{F}^{o}}{m}\right)\Theta\left(\omega+\frac{\hbar^{2}q^{2}}{2m}-\frac{\hbar^{2}q k_{F}^{o}}{m}\right)$ with $k_{F}^{o}=\sqrt{2 m \mu}/\hbar$. 

\begin{figure}[htbp]
\begin{center}
\includegraphics[width=85mm]{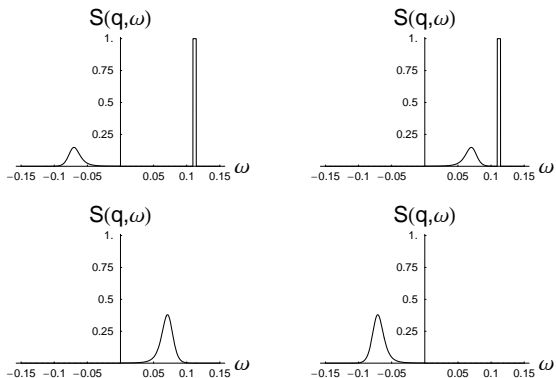}
\caption{Sonic Hawking radiation can be detected through the measurement of $S(q,\omega)$. 
The figure shows the $\omega$-dependence of $S(q,\omega)$ in the subsonic (left) and supersonic (right) regions and for a positive (top) and a negative (bottom) value of $q$. Here $k_{\text{B}}T_{\text{H}} = 0.15\, V_{\text{max}}$, $\epsilon_{\text{max}} = 2.5 \, V_{\text{max}}$ and $|q|=0.05\,m v_{\text{esc}}/\hbar$ is chosen in the region of small $q$ where thermal-like transitions (broadened peaks) dominate over the zero-temperature contributions (step-like peaks). $S(q,\omega)$ is in units of $L_{\text{eff}}\, m\,/\,2\pi\hbar^{3}q $ and $\omega$ is in units of $V_{\text{max}}/\hbar$. }
\label{figsqo}
\end{center}
\end{figure}

Due to the Pauli exclusion principle and to the one-dimensionality, which disconnects the Fermi-``surface'' into two points (represented by $v_{\text{L}}$ and $v_{\text{R}}$), $S(q,\omega)$ has a typical peaks structure, which contrasts the non-interacting two and three dimensional cases \cite{Pines}. The peaks are related to the dispersion relation $\omega(q)=-v_{\text{L}} \,|q|\,\Theta(-q)+v_{\text{R}}  \,|q|\,\Theta(q)$ valid for small $q$. In the supersonic region $\omega(q)$ is negative for $q<0$ meaning that small perturbations propagate always to the right.
Figure 2 shows the typical behaviour of $S(q,\omega)$.
The broad peaks in the bottom of Figure 2 ($q<0$) correspond to the negative-$q$ branch of the dispersion relation ($\omega(q)=-v_{\text{L}} \,|q|$). Also the broad peaks in the top of Figure 2 ($q>0$) correspond to the same dispersion relation but with opposite sign of $\omega$ and $q$, but representing these the inverse processes.
In contrast the sharp peaks in the top of Figure 2 correspond to the dispersion relation $\omega(q)=v_{\text{R}} \,|q|$, which is not broadened by the tunnelling through the barrier.
The characteristic wave-vectors of Hawking radiation is $q_{c}\sim k_{\text{B}}T_{\text{H}}/c\sim1/2\pi^{2}l_{x}^{2}n$ where $l_{x}=\sqrt{m/\hbar\omega_{x}}$ and $n$ is the density. Indeed
at very low frequencies $\hbar\omega \ll k_{\text{B}}T_{\text{H}}$ and momentum $S(q,\omega)$ is dominated by Hawking radiation. 
The peak values at $|\omega|=|v_{\text{L}}q| $ tends to $1/4$ of the corresponding zero-temperature value $L_{\text{eff}}\, m\,/\,2\pi\hbar^{3}q $ but with a half-width $\Delta\omega=1.76\, q k_{\text{B}}T_{\text{H}} / m v_{\text{L}}$, which is much larger than the corresponding half-width $\Delta\omega=\hbar q^{2} / 2 m $ of the zero-temperature contribution.

The experimental observation of the widths of these peaks would provide a measure of Hawking temperature. 
Degenerate Fermi atoms have been only recently obtained \cite{Lithium}.  
Lithium can have a degeneracy temperature as high as $8\;\mu$K, which fixes the scale of $\epsilon_{\text{max}}$.
 A reservoir of degenerate Lithium could be connected to non trapped  states with a 1D channel realized by optical potentials or wave-guides \cite{Chips}. 
The barrier potential could be created by a far-blue-detuned laser sheet propagating in a direction orthogonal to that of the flow propagation and tightly focused with a waist ($1/e^{2}$) of twice the laser wavelength ($760$ nm). Hawking temperature would be of order of $T_{\text{H}} \sim 200 $ nK and the characteristic wavelength of Hawking radiation $\lambda_{c}$ of order of $40\;\mu$m.

The dynamic structure factor $S(q,\omega)$ in the subsonic region is equivalent to that of a 1D channel connecting two reservoirs at temperatures $T=0$ on the left and $T=T_{\text{H}}$ on the right.
This is the analogue of Hawking radiation from a black hole into the surrounding zero-temperature space. 
To distinguish between this pure quantum state and a mixed thermal state it would require the measurement of correlations such as for instance those between opposite sides of the barrier.
Correlated pairs of quasi-particles (particle-hole), the ``real'' particles in an effective low-energy description, are created at the barrier and evolve in opposite directions.

The interpretation of the Hawking radiation as related to a tunnel effect is in agreement with a recent derivation of the Hawking radiation as a tunnelling process of particles in a dynamical geometry due to Parikh and Wilczek \cite{Wilczek00}. 
A relationship with a generalized form of diffraction has been pointed out by Sanchez \cite{Sanchez}.
Moreover electric pair production in a strong static electric field is related to a similar tunnelling through an inverted harmonic oscillator \cite{Heisenberg36}.
As quantum tunnelling is very sensitive to the details of the potential barrier, the distribution of radiated quasi-particles, related to the reflection coefficient of the barrier, can be different from (\ref{penetration}). 
In particular this suggest the possibility that some information on the supersonic part of the fluid (the analogue of the classical forbidden region of a black hole) is carried by the radiation. 

\emph{Part II. 1D quantum fluids.} --- 
Hawking's prediction is, in principle, valid also for strongly correlated 1D systems, such as the Tonks and Lieb-Liniger gases \cite{TonksGas}, which are currently subject of intense interest \cite{Goerlitz01} and electrons in clean short quantum wires \cite{Glazman88,Glazman93}.
Using standard hydrodynamics of 1D flow, it is possible to rewrite Hawking temperature (\ref{unruh}) in a more general form \cite{notehydro} as
$ T_{\text{H}}=\eta \, \hbar \omega_{x} / 2\pi k_{\text{B}} $ 
where $\eta$ is defined by
\begin{equation}
\eta =\sqrt{\frac{3}{4}+\frac{1}{4}\left(n \frac{d^2 h }{  dn^2} / \frac{d h }{ dn} \right)}\;.
\label{prefactor}
\end{equation}
In general $\eta$ is only weakly dependent on the form of the function enthalpy $h(n)$ and Hawking temperature is basically proportional to a quantity that is related to the propagation of a single particle of the fluid in the vicinity of the horizon.
In the case of a non-interacting degenerate 1D Fermi gas considered here, $\eta$ is equal to $1$ since the enthalpy is proportional to the square of the density. In the case of a 1D Bose gas, $\eta$ is equal to $\sqrt{3/4}$ in the mean-field regime \cite{Giovanazzi04} since the enthalpy is linear in the density and is equal to $1$ in the Tonks-Girardeau limit \cite{TonksGas} since the enthalpy is identical to that of a Fermi-degenerate gas in that limit. 
Beside the importance of testing some aspects of black hole evaporation, 
sonic Hawking radiation would be a remarkable transport phenomena in these strongly correlated quantum fluids and could be used to investigate their properties.

A quite different 1D system where one may consider to experimentally realize a sonic black hole and to test Hawking radiation would be a system of charge particles, like electrons in quantum wires or ions in elongated traps. 
The long-range part of the Coulomb interaction could substantially renormalize the Hawking temperature in long wires in a way similar to how it does for the scattering from an impurity \cite{Glazman93}.

\emph{Conclusion} --- A Fermi-degenerate flow of particles propagating in a 1D channel in the presence of a smooth effective barrier provides an exact solvable quantum mechanical model for the sonic analogue of Hawking radiation. The agreement with Hawking's calculation is obtained in the limit of a very smooth barrier, which is the equivalent in Hawking's derivation to a slowly evolving event horizon.
The thermal and incoherent characteristic of Hawking radiation is discussed in terms of a measurable quantity such as the dynamic structure factor.

I thank U. Leonhardt and T. Kiss for inspiring discussions on sonic black holes in atomic Bose-Einstein condensates. I acknowledge the Marie Curie Programme of the European Commission for funding.


\begin{thebibliography}{15}
\bibitem{eff} 
{$V_{\text{ext}}(x)= \langle\xi_{0;x}| -(\hbar^{2}/2m)(\partial_{y}^{2}+\partial_{z}^{2})+ V(x,y,z)|\xi_{0;x} \rangle$ where $V(x,y,z)$ is the external potential.}
\bibitem{notecurrent}
{The current conservation can be written in a form that is reminescent of the quantization of the conductance in 1D channels: $\Omega = (\mu_R - \mu_L)/ h_{\text{P}}$ where $h_{\text{P}}$ is the Planck constant, $\mu_{L}= m v_{\text{esc}}^{2}/2$ and $\mu_{R}= m v_{\text{max}}^{2}/2 $.}
\bibitem{note}
{The energy-weighted sum rule $m_{1}=\int_{-\infty}^{\infty}S(q,\omega)\,\hbar\omega \,d\omega$ 
evaluated with the Fermi fluid at rest gives 
the generalized form of the f-sum rule $m_{1} = N_\text{eff}(\hbar^{2}q^{2}/2m)$ where $N_\text{eff}$ is an effective number of atoms given by $N_\text{eff}=(k_{F}/\pi)\sqrt{\pi/2}L_{p} $. Here $\sqrt{\pi/2}L_{p} $ is the effective length of the region of interaction between the flow and the probe. A similar consideration holds for the compressibility sum rule at small $q$, i.e. $m_{-1} = \int_{-\infty}^{\infty}S(q,\omega)\,(1/\hbar\omega) \,d\omega= N_\text{eff}(1/2m\hbar^{2}v_{F}^{2})$. The f-sum rule holds for the fluid in motion only after averaging between positive and negative values of $q$.}
\bibitem{notehydro}
{The lhs of the Bernoulli equation rewritten as $\Omega^2/ 2 n^2  +  h(n) = \mu/m - V_{\text{ext}} (x)/ m$ where  $\Omega = n v$ is the conserved current has a minimum at a finite value of $n$
that corresponds to the transition between subsonic and supersonic flow.
The expansion of $V_{\text{ext}}$ around its maximum fixes the value of Hawking temperature (\ref{unruh}).}
\end{thebibliography}
\end{document}